\documentstyle[12pt]{article}

\catcode`\@=11
\long\def\@makefntext#1{
\protect\noindent \hbox to 3.2pt {\hskip-.9pt
$^{{\ninerm\@thefnmark}}$\hfil}#1\hfill}		

 \def\@makefnmark{\hbox to 0pt{$^{\@thefnmark}$\hss}}  

\def\ps@myheadings{\let\@mkboth\@gobbletwo
\def\@oddhead{\hbox{}
\rightmark\hfil\ninerm\thepage}
\def\@oddfoot{}\def\@evenhead{\ninerm\thepage\hfil
\leftmark\hbox{}}\def\@evenfoot{}
\def\sectionmark##1{}\def\subsectionmark##1{}}

\newcounter{sectionc}\newcounter{subsectionc}\newcounter{subsubsectionc}
\renewcommand{\section}[1] {\vspace{0.6cm}\addtocounter{sectionc}{1}
\setcounter{subsectionc}{0}\setcounter{subsubsectionc}{0}\noindent
	{\bf\thesectionc. #1}\par\vspace{0.4cm}}
\renewcommand{\subsection}[1] {\vspace{0.6cm}\addtocounter{subsectionc}{1}
	\setcounter{subsubsectionc}{0}\noindent
	{\it\thesectionc.\thesubsectionc. #1}\par\vspace{0.4cm}}
\renewcommand{\subsubsection}[1]
{\vspace{0.6cm}\addtocounter{subsubsectionc}{1}
	\noindent {\rm\thesectionc.\thesubsectionc.\thesubsubsectionc.
	#1}\par\vspace{0.4cm}}

\newcounter{appendixc}
\newcounter{subappendixc}[appendixc]
\newcounter{subsubappendixc}[subappendixc]

\renewcommand{\appendix}[1] {\vspace{0.6cm}
        \refstepcounter{appendixc}
        \setcounter{figure}{0}
        \setcounter{table}{0}
        \setcounter{equation}{0}
        \renewcommand{\thefigure}{\Alph{appendixc}.\arabic{figure}}
        \renewcommand{\thetable}{\Alph{appendixc}.\arabic{table}}
        \renewcommand{\theappendixc}{\Alph{appendixc}}
        \renewcommand{\theequation}{\Alph{appendixc}.\arabic{equation}}
        \noindent{\bf Appendix \theappendixc #1}\par\vspace{0.4cm}}

\def\abstracts#1{{
	\centering{\begin{minipage}{30pc}\tenrm\baselineskip=12pt\noindent
	\centerline{\tenrm ABSTRACT}\vspace{0.3cm}
	\parindent=0pt #1
	\end{minipage}}\par}}

\renewenvironment{thebibliography}[1]
	{\begin{list}{\arabic{enumi}.}
	{\usecounter{enumi}\setlength{\parsep}{0pt}
\setlength{\leftmargin 1.25cm}{\rightmargin 0pt}
	 \setlength{\itemsep}{0pt} \settowidth
	{\labelwidth}{#1.}\sloppy}}{\end{list}}

\newcounter{itemlistc}
\newcounter{romanlistc}
\newcounter{alphlistc}
\newcounter{arabiclistc}

\newcommand{\fcaption}[1]{
        \refstepcounter{figure}
        \setbox\@tempboxa = \hbox{\tenrm Fig.~\thefigure. #1}
        \ifdim \wd\@tempboxa > 6in
           {\begin{center}
        \parbox{6in}{\tenrm\baselineskip=12pt Fig.~\thefigure. #1}
            \end{center}}
        \else
             {\begin{center}
             {\tenrm Fig.~\thefigure. #1}
              \end{center}}
        \fi}

\newcommand{\tcaption}[1]{
        \refstepcounter{table}
        \setbox\@tempboxa = \hbox{\tenrm Table~\thetable. #1}
        \ifdim \wd\@tempboxa > 6in
           {\begin{center}
        \parbox{6in}{\tenrm\baselineskip=12pt Table~\thetable. #1}
            \end{center}}
        \else
             {\begin{center}
             {\tenrm Table~\thetable. #1}
              \end{center}}
        \fi}

\def\@citex[#1]#2{\if@filesw\immediate\write\@auxout
	{\string\citation{#2}}\fi
\def\@citea{}\@cite{\@for\@citeb:=#2\do
	{\@citea\def\@citea{,}\@ifundefined
	{b@\@citeb}{{\bf ?}\@warning
	{Citation `\@citeb' on page \thepage \space undefined}}
	{\csname b@\@citeb\endcsname}}}{#1}}

\newif\if@cghi
\def\cite{\@cghitrue\@ifnextchar [{\@tempswatrue
	\@citex}{\@tempswafalse\@citex[]}}
\def\citelow{\@cghifalse\@ifnextchar [{\@tempswatrue
	\@citex}{\@tempswafalse\@citex[]}}
\def\@cite#1#2{{$\null^{#1}$\if@tempswa\typeout
	{IJCGA warning: optional citation argument
	ignored: `#2'} \fi}}

\def\fnt#1#2{\footnotetext{\kern-.3em
	{$^{\mbox{\sevenrm #1}}$}{#2}}}

 1
 1
 1

\font\tenbf=cmbx10
\font\tenrm=cmr10
\font\tenit=cmti10

\font\ninerm=cmr9

\newcommand{\be}{\begin{equation}}
\newcommand{\ee}{\end{equation}}

\newcommand{\cem}{\hspace{1cm}}

\newcommand{\al}{{\alpha}}

\newcommand{\da}{{\dot{\alpha}}}

\newcommand{\phib}{{\bar{\phi}}}
\newcommand{\Fb}{{\bar{F}}}

\newcommand{\ca}{{\cal A}}
\newcommand{\cd}{{\cal D}}
\newcommand{\cf}{{\cal F}}
\newcommand{\cl}{{\cal L}}
\newcommand{\cq}{{\cal Q}}
\newcommand{\cw}{{\cal W}}

\newcommand{\A}{\mbox{A}}
\newcommand{\tr}{\mbox{tr}}
\begin{document}

\centerline{\tenbf SUPERGRAVITY, LINEAR MULTIPLETS,}
\baselineskip=16pt
\centerline{\tenbf AND CHERN-SIMONS FORMS}
\vspace{0.8cm}
\centerline{\tenrm R. GRIMM}
\baselineskip=13pt
\centerline{
\tenit Centre de Physique Th\'eorique - CNRS - Luminy, Case 907}
\baselineskip=12pt
\centerline{\tenit F-13288 Marseille Cedex 9, France}
\vspace{2.4cm}
\abstracts{Some general features of locally supersymmetric
theories (N=1 in four dimensions) involving Chern-Simons
forms and antisymmetric tensors are sketched out.
The relevance of the three-form multiplet both for the
description of Chern-Simons forms and the supersymmetry
properties of the gaugino condensate is pointed out.}

\vspace{2.0cm}
\rm\baselineskip=14pt
\section{Introduction}
The supersymmetric standard model has a rather simple structure
concerning the basic building blocks involved in its construction:
only chiral and Yang-Mills multiplets are needed.
On the other hand, when it comes to the discussion of
low energy effective theories of superstrings one has to
deal with more sophisticated ingredients.
First of all the general K\"ahler structure, in particular the
K\"ahler phase transformations, should be taken into account.
This can be done in $U_K(1)$ superspace\cite{BGGM1,BGG}, a geometric
framework which embodies at the same time supergravity and matter,
with K\"ahler transformations appearing ab initio in the structural
group, on the same footing as the local Lorentz transformations
of supergravity.

Moreover, in superstring inspired scenarios, less familiar
multiplet structures and new couplings appear, as for instance
linear multiplets and Chern-Simons forms induced from
the Green-Schwarz anomaly cancellation
in the superstring\cite{FV,CFV}.
Again, $U_K(1)$ superspace provides an appropriate background
for a geometric description of these new structures\cite{BGGM2,ABGG}.
In turn, supersymmetric Chern-Simons forms are closely related
to the three-form multiplet\cite{Gates}: the three-form superspace
geometry provides the relevant framework for the Chern-Simons forms,
Yang-Mills as well as gravitational\cite{GG1,GG2}.
In a similar line of reasoning it will be pointed out here
that the three-form multiplet is the relevant object which accounts
for the supersymmetry properties of the gaugino condensate
supermultiplet. Finally, it might be conceivable that three-form
multiplets enter the stage of supersymmetric theories in their own
right.

\section{$U_K(1)$ Superspace: Unification of Supergravity and Matter}
The basic object in this geometric formulation is the frame
$E^A = dz^M E_M{}^A(z)$, a differential one-form in superspace,
which, as an extension of traditional general relativity,
introduces the usual moving frame $e_m{}^a(x)$ together with
the Rarita-Schwinger field $\psi_m{}^\al$, its
supersymmetry partner, in a unified manner. In distinction
to the traditional approach\cite{WB}, the frame is not only
subject to local Lorentz transformations, but also to local chiral
transformations, with chiral weights defined as
\be
w(E^a) \ = \ 0, \cem w(E^\al) \ = \ +1, \cem w(E_\da) \ = \ -1.
\ee
Correspondingly, in the definition of the torsion a new term appears,
in addition to the familiar spin connection:
\be T^A \ = \ d E^A + E^B \Phi_B{}^A + w(E^A) E^A \A. \ee

The crucial point is that the $U_K(1)$ gauge potential $\A = E^A \A_A$
is given in terms of the K\"ahler potential $K(\phi,\phib)$
of the matter sector - the traditional unconstrained
prepotential of supersymmetric gauge theory has been
consistently replaced by the K\"ahler potential, which is a function
of the chiral and antichiral matter superfields, giving rise
to a unified geometric description of the supergravity-matter system.
In this construction K\"ahler transformations are built in from
the very beginning, on the same footing as local Lorentz
transformations. The transformation
\be K(\phi,\phib) \ \mapsto \ K(\phi,\phib) + F(\phi) + \Fb(\phib), \ee
of the K\"ahler potential itself induces, by construction\cite{BGG},
a transformation
\be \A \ \mapsto \ \A + \frac{i}{2} d \, Im \, F, \ee
on the $U_K(1)$ gauge potential which in turn serves to covariantize
the K\"ahler phase transformations
\be
E^A \ \mapsto \ E^A \exp \left(-\frac{i}{2} w(E^A) Im \, F \right),
\ee
of the frame in superspace. Observe that these assignements determine
completely the K\"ahler properties of the supergravity superfields.
This holds in particular for the superfields $R$, $R^\dagger$ and
$G_a$ of weights $w(R)=+2$, $w(R^\dagger)=-2$ and $w(G_a)=0$,
which are subject to the conditions
\be \cd^\da R \ = \ 0, \cem  \cd_\al R^\dagger \ = \ 0, \ee
and
\be
\cd^\al \cd_\al R - \cd_\da \cd^\da R^\dagger \ = \ 4i \cd_a G^a,
\ee
with Lorentz and K\"ahler covariant derivatives.
For a more exhaustive presentation of $U_K(1)$ superspace see the
literature cited so far.

\section{Yang-Mills in $U_K(1)$ Superspace and Chern-Simons Forms}
Having set up the geometrical framework for the supergravity-matter
system we shall now include gauge interactions. To this end we consider
matter fields $\phi$ and $\phib$ in some representation of the
gauge group and with gauge transformations
\be
{}^g \phi \ = \ g^{-1} \phi, \cem {}^g \phib \ = \ \phib g,
\ee
where $g$ is a superspace dependent group element, like
$g(z)=\exp \, i \al^{(r)} (z) T_{(r)}$, with superfield gauge parameters
$\al^{(r)} (z)$ and generators $T_{(r)}$. Exterior covariant derivatives
\be
d \phi - \ca \phi \ = \ D \phi \ = \ E^A \cd_A \phi, \cem
d \phib + \phib \ca \ = \ D \phib \ = \ E^A \cd_A \phib,
\ee
are defined in terms of the superspace one-form Lie algebra valued
gauge potential
\be
\ca \ = \ E^A \ca_A^{(r)} T_{(r)}, \cem
{}^g \ca \ = \ g^{-1} \ca g - g^{-1} d g,
\ee
and the matter fields are now constrained to be covariantly chiral,
resp. antichiral:
\be \cd^\da \phi \ = \ 0, \cem \cd_\al \phi \ = \ 0. \ee

As is well known, the covariant field strength
\be d \ca + \ca \ca \ = \ \cf \ = \ \frac{1}{2} E^A E^B \cf_{BA}, \ee
is subject to constraints.
Correspondingly, the gaugino superfields $\cw_\al$ and $\cw^\da$
of K\"ahler weights $w(\cw_\al)=+1$ and $w(\cw^\da)=-1$,
respectively, are subject to
\be \cd_\al \cw^\da \ = \ 0, \cem \cd^\da \cw_\al \ = \ 0, \ee
\be \cd^\al \cw_\al \ = \ \cd_\da \cw^\da. \ee
In turn, as a consequence of this constrained superspace geometry,
the gaugino-squared superfield $\tr \, \cw^2$ and its complex
conjugate are chiral, resp. antichiral superfields with the
additional property
\be
\left( \cd^\al \cd_\al - 24 R^\dagger \right) \tr \, \cw^2
- \left( \cd_\da \cd^\da - 24 R \right) \tr \, \overline{\cw}^2
\ = \ -2i \varepsilon^{dcba} \tr \left( \cf_{dc} \cf_{ba} \right).
\ee

This last equation has a very natural interpretation in relation
with supersymmetric Chern-Simons forms built from the superspace
gauge potential $\ca$ defined as
\be \tr \left( \ca d \ca + \frac{2}{3} \ca \ca \ca \right)
\ = \ \cq  \ = \ \frac{1}{3} E^A E^B E^C \cq_{CBA}, \ee
and satisfying
$d \cq \ = \ \tr \left( \cf \cf \right)$.
We will come back to this issue in a short while after having
introduced the linear multiplet and its superspace geometry.

\section{Superfield Actions}
Given the geometric structures of the previous two sections,
we turn now to the invariant actions. In the geometric approach,
invariance under general coordinate transformations and under
supersymmetry means nothing else than invariance under
general reparametrizations in superspace. In addition, the
actions should be invariant under Lorentz, K\"ahler and Yang-Mills
transformations. It turns out that already the simplest
invariant, namely the basic volume density, is quite non-trivial:
it provides at the same time the kinetic terms for
supergravity and for the matter sector:
\be \cl_{supergravity \, + \, matter} \ = \ -3 \int E. \ee
Here the integration is performed over full superspace, the
commuting and the anticommuting directions. In the kinetic
terms for the Yang-Mills sector,
\be \cl_{Yang-Mills} \ = \
\frac{1}{8} \int \frac{E}{R}
\, f_{(r)(s)}(\phi) \, \cw^{\al (r)} \cw_\al^{(s)} +
\frac{1}{8} \int \frac{E}{R^\dagger}
\, \bar{f}_{(r)(s)}(\phib) \, \cw_\da^{(r)} \cw^{\da (s)},
\ee
chiral and antichiral volume elements of non-trivial
K\"ahler weights are used. In general,
holomorphic gauge coupling functions can be present.
In the potential terms,
\be \cl_{pot} \ = \
\frac{1}{2} \int \frac{E}{R} e^{K/2} W(\phi) +
\frac{1}{2} \int \frac{E}{R^\dagger} e^{K/2} \bar{W}(\phib),
\ee
the superpotential transforms as
\be
W(\phi) \ \mapsto \ e^{-F(\phi)} W(\phi), \cem
\bar{W}(\phib) \ \mapsto \ e^{-\bar{F}(\phib)} \bar{W}(\phib),
\ee
and K\"ahler invariance is established due to the factors $e^{K/2}$.
In other words, the combinations
\be e^{K/2} W(\phi), \cem e^{K/2} \bar{W}(\phib), \ee
have well-defined K\"ahler weights
\be w \left( e^{K/2} W(\phi) \right) \ = \ +2, \cem
w \left( e^{K/2} \bar{W}(\phib) \right) \ = \ -2, \ee
which are compensated by those of $R$ and $R^\dagger$, respectively.

The sum of these three separately invariant actions provides
the dynamics of the complete traditional supergravity-matter
and Yang-Mills system. In its explicit expansion in terms of
component fields, which still requires some computational and
book-keeping effort\cite{BGG}, the combinations
\be
\cd^\al \cd_\al R + \cd_\da \cd^\da R^\dagger,
\ee
\be
\cd^\al \cd_\al \, \tr \, \cw^2 + \cd_\da \cd^\da \, \tr \, \overline{\cw}^2,
\ee
orthogonal to those of Eq. (7) and Eq. (15), are of crucial
importance.

\section{Linear Multiplet and Chern-Simons Forms}
The linear multiplet is another supermultiplet of helicity content
$(0,1/2)$. In distinction to the chiral multiplet which is
composed of a complex scalar, a Majorana spinor and a complex
scalar auxiliary field, the linear multiplet contains an
antisymmetric tensor gauge potential, a real scalar and a Majorana
spinor, no auxiliary field. The presence of the antisymmetric tensor
\be b_{mn} \ = \ - b_{nm}, \cem
b_{mn} \ \mapsto \ b_{mn} + \partial_m \xi_n - \partial_n \xi_m, \ee
makes the linear multiplet a gauge multiplet. As an antisymmetric
tensor should be viewed as a two-form gauge potential, the linear
multiplet has a well-established geometric formulation
based on the superspace two-form
\be B \ = \ \frac{1}{2} E^A E^B B_{BA} (z). \ee
As the corresponding field-strength, obtained from applying the
exterior derivative, is a three-form, it may be combined with
the Chern-Simons form $\cq$ encountered before in the Yang-Mills case,
such that
\be d B + k \cq \ = \ H \ = \ \frac{1}{3!} E^A E^B E^C H_{CBA}, \ee
with some constant $k$. The Bianchi identity is then simply
\be d H \ = \ k \, \tr (\cf \cf). \ee

At this point a detailed anlysis of the superspace structure reveals
that this provides indeed a compatible geometric system and the
Bianchi identities correspond to the modified linearity conditions
\be
\left( \cd^2 - 8 R^\dagger \right) L \ = \ 2k \, \tr \, \overline{\cw}^2,
\ee
\be
\left( \overline{\cd}^2 - 8 R \right) L \ = \ 2k \, \tr \, \cw^2.
\ee
These equations characterize the properties of the linear multiplet,
coupled to Chern-Simons forms, in analogy to the chirality constraints
of the usual matter superfields. They will become crucial in the
evaluation of component field actions and the discussion of the
ensuing interaction terms. Another information from the superspace
Bianchi identities is the equation
\be
\left( [\cd_\al, \cd_\da] - 4 \sigma_{\al \da}^a \, G_a \right) L
\ = \ - \frac{1}{3} \sigma_{d \al \da} \varepsilon^{dcba} H_{cba}
- 4k \, \tr \left( \cw_\al \cw_\da \right),
\ee
which identifies the field strength tensor $H_{cba}$
in the superfield expansion of the linear superfield.

\section{Invariant Actions with Linear Multiplets}
The coupling of the linear multiplet (for simplicity, only one
linear multiplet is considered here, in general an arbitrary
number is possible) can be incorporated in promoting
the superfield K\"ahler potential, which so far was a function of the
chiral and antichiral matter superfields only, to a more general
function
\be K(\phi, \phib, L), \ee
depending on the linear superfield $L$ as well. In this case, the
action obtained from the volume of superspace, Eq. (17), will now
describe the kinetic terms for the linear multiplet as well. In
addition, due to the modified linearity conditions, {\it i.e.} the
gaugino superfield squared terms, it will also give rise to gauge kinetic
terms with a gauge coupling function proportional to
$\partial K/ \partial L$. Hence, the simple volume term of superspace,
when expanded in terms of component fields reveals quite a lot of
information.

On the other hand, in the explicit evaluation of the component field action
one finds that the normalization of the curvature scalar acquires a field
dependend contribution proportional to
\be \left( 1 - \frac{1}{3} L \frac{\partial K}{\partial L} \right)^{-1}. \ee
In the geometric approach used here, the normalization of the Einstein term
can easily be modified making use of the super-Weyl transformations
adopted to $U_K(1)$ superspace with linear superfield (whose proper
rescalings must be taken into account too). Parametrizing the rescaling
in terms of a superfield $F(\phi, \phib, L)$ (not to be confused with
K\"ahler transformations denoted similarly), and requiring, for instance,
canonical normalization for the curvature scalar, one finds that
the corresponding action is given as
\be \cl_{supergravity \, + \, chiral \ matter \, + \, linear \, + \,
Yang-Mills}
 \ = \ -3 \int E \, F(\phi, \phib, L), \ee
with the function $F$ related to the modified K\"ahler potential
through the equation\cite{BGGM2,nonhol,ABGG}
\be F - L \frac{\partial F}{\partial L} \ = \
1 - \frac{1}{3} L \frac{\partial K}{\partial L}, \ee
implying that $F$ should be of the general form
\be F(\phi, \phib, L) \ = \ 1 + L \, V(\phi,\phib)
      + \frac{1}{3} L \int \frac{dL}{L}\frac{\partial K}{\partial L}. \ee
Here, $V(\phi,\phib)$ is a new arbitrary function, relevant for
supersymmetric theories with non-holomorphic gauge coupling
functions\cite{Jan}.
Also, it is precisely this superfield which appears in effective
theories with K\"ahler anomaly cancellation
mechanism\cite{CO,DFKZ,nonhol,ABGG}.
The superfield appearance of the explicit Yang-Mills action, Eq. (18),
and of $\cl_{pot}$, Eq. (19), remain unchanged.

\section{Linear Multiplet vs. Chiral Multiplet}
The modified linearity conditions, Eqs. (29,30), can be obtained from
a variational principle in superspace. This goes as follows. First
of all we express the gaugino-squared terms as derivatives of the
Chern-Simons superfield,
\be
\tr \, \overline{\cw}^2 \ = \
\frac{1}{2} \left( \cd^2 - 8 R^\dagger \right) \Omega,
\ee
\be
\tr \, \cw^2 \ = \
\frac{1}{2} \left( \overline{\cd}^2 - 8 R \right) \Omega.
\ee
Then one writes down the first order action
\be \cl_{(1)} \ = \ -3 \int \left(
F(\phi, \phib, L) + (L-k \Omega)(S+\bar{S}) \right). \ee
In this action $L$ is understood to be unconstrained, while
$S$ and $\bar{S}$ are chiral resp. antichiral. Variation
of this action with respect to those chiral superfields,
taking into account the chirality constraints and integration
by parts in superspace, results in the modified linearity
equations (29,30) and we get back the theory
described in the previous section. On the other hand,
variation of the first order action with respect to $L$
yields\cite{nonhol}
\be (S+\bar{S})
\left( 1 - \frac{1}{3} L \frac{\partial K}{\partial L} \right)
\ = \ \frac{1}{3} F \frac{\partial K}{\partial L}
- \frac{\partial F}{\partial L}. \ee
This equation should be read as an equation which serves to express
$L$ in terms of $\phi$, $\phib$ and of $S+\bar{S}$, giving rise
to a theory entirely in terms of chiral multiplets with the special
property that $S$ and $\bar{S}$ appear only through their sum.

One should, however, again keep track of the normalization of
the curvature scalar: canonical normalization is established if
one requires the condition
\be F(\phi, \phib, L) + L (S+\bar{S}) \ = \ 1, \ee
where $L$ is solution of the above equation.
Combination of these two equations leads again to
Eq. (35) of the previous section.

It is in this sense that a linear multiplet is said to be dual to
a chiral multiplet. In this dual theory, $K$ is a true K\"ahler potential,
it depends only on chiral superfields, and one might use this
correspondance to give a geometric interpretation to the function
$K(\phi,\phib,L)$ in the previous theory.
As an example consider\cite{nonhol}
\be K(\phi,\phib,L) \ = \ K_0(\phi,\phib) + \al \log L, \ee
\be F(\phi,\phib,L) \ = \ 1 - \frac{1}{3} \al + L \, V(\phi,\phib). \ee
Eq. (40) is solved by $\al / 3L = S + \bar{S} + V(\phi, \phib)$
yielding
\be K(\phi,\phib,S + \bar{S}) \ = \ K_0(\phi,\phib) + \al \log \al / 3
- \al \log \left(S + \bar{S} + V(\phi, \phib) \right). \ee

\section{Three-Form Multiplet}
The three-form multiplet is yet another helicity $(0,1/2)$
supermultiplet. It consists of a complex scalar, a Majorana spinor,
an antisymmetric three-index tensor gauge potential and a real
scalar auxiliary field. Again, it is a gauge multiplet, and its
properties can be determined in superspace geometry\cite{Gates}.
As a starting point consider
\be C \ = \ \frac{1}{3!} E^A E^B E^C C_{CBA} (z), \ee
a three-form in $U_K(1)$ superspace with field strength
\be dC \ = \ \Sigma \ = \
\frac{1}{4!} E^A E^B E^C E^D \Sigma_{DCBA}, \ee
and Bianchi identity $d \Sigma = 0$. A detailed superspace
analysis allows to identify the three form multiplet
in terms of a chiral superfield $T$,
\be \cd^\da T \ = \ 0, \cem \cd_\al \bar{T} \ = \ 0, \ee
with an additional constraint
\be
\left( \cd^\al \cd_\al - 24 R^\dagger \right) T
- \left( \cd_\da \cd^\da - 24 R \right) \bar{T}
\ = \ \frac{8i}{3} \varepsilon^{dcba} \Sigma_{dcba},
\ee
where the vectorial fieldstrength superfield
\be \varepsilon^{dcba} \Sigma_{dcba} \ = \
\varepsilon^{dcba} \left( 4 \, \cd_d C_{cba}
+6 \, T_{dc}{}^\al C_{\al \, ba} +6 \, T_{dc \, \da} C^\da{}_{ba} \right), \ee
appears. The solution of these constraints is given in terms of an
unconstrained prepotential superfield $\Omega$ such that
\be \bar{T} \ = \ \left( \cd^2 - 8 R^\dagger \right) \Omega, \cem
       T \ = \ \left( \overline{\cd}^2 - 8 R \right) \Omega. \ee
As we are working in $U_K(1)$ superspace the K\"ahler chiral
weights are read off to be $w(T)=+2$ and $w(\bar{T})=-2$. In
coupling the three-form multiplet one may perfectly well
include the superfields $T$ and $\bar{T}$ into the K\"ahler
potential and the function $F$, with appropriate care, however,
to the normalization of the Einstein term.
As to the superpotential, one should maintain the
K\"ahler transformations defined in Eq. (20).
Expanding the superpotential $\cw(\phi,T)$
in powers of $T$, this may be achieved
with suitable insertions of exponentials of the K\"ahler potential
\be \cw(\phi,T) \ = \
\sum_{n \geq 0} W_n(\phi) \left( e^{-K/2}T \right)^n, \ee
and K\"ahler transformations
\be  W_n \ \mapsto \ e^{(n-1)F} W_n. \ee
of the coefficient functions, thus establishing
$\cw \ \mapsto \ e^{-F} \cw$.

\section{Comments}
The investigation of the three-form multiplet is less
academic as it might appear. First of all it seems that
this multiplet structure applies to the description of
the supercurrent multiplet and its anomaly structure.
On the other hand it has been advantageously employed
in the study of supersymmetric Chern-Simons forms,
both Yang-Mills and gravitational\cite{GG1,GG2}:
four-dimensional Chern-Simons forms are three-forms
which change under gauge-transformations by the
the exterior derivative of a two-form.

In a similar line of reasoning the three-form multiplet
should be relevant in gluino condensation: if the
supersymmetry properties of the condensate are supposed
to reflect those of the superfield $tr \, \cw^2$ and its
complex conjugate, the natural multiplet structure for
an effective description of the gluino condensate
is the three-form. This is related to the fact that
the gaugino superfield is not just chiral, Eq. (13), but
subject to the additional condition Eq.(14). As a consequence,
the superfield $tr \, \cw^2$ is not just chiral either,
but subject to the same additional restrictions as the
three-form (cf. Eq. (15) and Eq. (48)).

Apart from this possibility of parametrizing the gaugino
condensate, the couplings of generic three-form multiplets
may of course be investigated in their own right.
Let me close with the description of a mechanism
paraphrasing the Chern-Simons - antisymmetric tensor coupling
on the level of the three form. To this end consider
abelian one- and two-form gauge potentials $A$ and $B$ with
fieldstrength $F=dA$ and $H=dB$. In addition, consider the
three form gauge potential $C$ with field strength $\Sigma$
defined as
\be \Sigma \ = \ d C + \tau H \, A. \ee
Inspection of this geometric structure in superspace shows
that the one-, two- and three-form geometries are indeed
compatible with this definition of $\Sigma$. Whereas in the one-
and two-form sectors one has the usual equations for the
gaugino and linear superfields, the chiral, antichiral superfields
$T$, $\bar{T}$ are now subject to a modified additional constraint
\begin{eqnarray} \lefteqn{
\left( \cd^\al \cd_\al - 24 R^\dagger \right) T
- \left( \cd_\da \cd^\da - 24 R \right) \bar{T}
- \frac{8i}{3} \varepsilon^{dcba} \Sigma_{dcba} } \nonumber \\
& &+ 16 \tau \left( \cd^\al \cw_\al + \cd_\da \cw^\da \right) L
+ 64 \tau \left( \cw^\al \cd_\al L + \cw_\da \cd^\da L \right) \ = \ 0,
\end{eqnarray}
reflecting the properties of the modified Bianchi-identities
$d\Sigma = \tau H \, F$. It is then a straightforward task to
implement this geometrical structure in a supersymmetric
dynamical context, which will be presented in a forthcoming
publication\cite{really?}.

\section{Acknowledgements}
The material presented here is based on work carried out over
the years in collaboration with P. Adamietz, P. Bin\'etruy,
G. Girardi and M. M\"uller.

\section{References}
This list of references is by no means intended to be exhaustive,
I apologize in advance for any undue omission.

\end{document}